\documentclass[submission,Phys]{SciPost}
\usepackage{authblk}

\usepackage[utf8]{inputenc} % input encodings (allow UTF-8 input)
\usepackage[T1]{fontenc} 	% selecting font encodings (use 8-bit T1 fonts)
\usepackage[english]{babel} % multilingual support (English language/hyphenation)

\usepackage[bitstream-charter]{mathdesign}
\usepackage{geometry} 		% flexible and complete interface to document dimensions
\usepackage{amsmath} 		% math package (American Mathematical Society)
\usepackage{amsthm} 		% math package (typesetting theorems using AMS style)
\usepackage{mathtools} 		% math package (fixes various deficiencies of amsmath)
\usepackage{float} 			% floating objects such as figures and tables
\usepackage{graphicx} 		% enhanced support for graphics
\usepackage{tabularx} 		% tabulars with adjustable-width columns
\usepackage{booktabs} 		% professional-quality tables
\usepackage{color, xcolor} 	% foreground and background colour management
\usepackage{pdfpages} 		% inclusion of external multi-page PDF documents
\usepackage{extarrows} 		% extra arrows beyond those provided in amsmath
\usepackage{multirow} 		% create tabular cells spanning multiple rows
\usepackage{multicol} 		% intermix single and multiple columns
\usepackage{caption} 		% customising captions in floating environments
\usepackage{subcaption} 	% support for sub-captions (captions for subfigures)
\usepackage{enumitem} 		% control layout of itemize, enumerate, description
\usepackage{setspace} 		% set space between lines (single, onehalf, double)
\usepackage{xspace} 		% define commands that appear not to eat spaces
\usepackage{ragged2e} 		% alternative versions of ragged-type commands
\usepackage{stackrel} 		% enhancement to the \stackrel command
\usepackage{tikz} 			% create PostScript and PDF graphics
\usepackage{braket} 		% Dirac bra-ket notation
\usepackage{bm} 			% access bold symbols in maths mode
\usepackage{tensor} 		% typeset tensors (tensor-style super- and subscripts)
\usepackage{slashed} 		% slash through characters (Feynman slash notation)
\usepackage{siunitx} 		% SI units package (typesetting values with units)
\usepackage{lastpage} 		% reference last page
\usepackage{cite} 			% improved citation handling
\usepackage[normalem]{ulem} % package for underlining
\usepackage{fontawesome} 	% access to web-related icons
\usepackage{tocloft} 		% control over the typography of the TOC, LOF, and LOT
\usepackage{titlesec} 		% interface to sectioning commands (various title styles)
\usepackage{doi} 			% create correct hyperlinks for DOI numbers
\usepackage{hyperref} 		% hypertext links (handel cross-referencing commands)
\usepackage[most]{tcolorbox} 					% coloured and framed text boxes
\usepackage[nameinlink, capitalize]{cleveref} 	% intelligent cross-referencing
\usepackage[nottoc, notlot, notlof]{tocbibind} 	% add references/index/contents to TOC
\usepackage[ruled, vlined]{algorithm2e} 		% floating algorithm environment
\hypersetup{
	pdftitle={Machine Learning and LHC Event Generation},
	pdfauthor={Butter et al.},
	colorlinks=true, 			% false: boxed links, true: colored links
	linkcolor={red!50!black}, 	% color of internal links (sections, pages, etc.)
	citecolor={blue!50!black}, 	% color of citation links (links to bibliography)
	urlcolor={blue!80!black} 	% color of URL links (external links)
} 
\DeclareSymbolFont{usualmathcal}{OMS}{cmsy}{m}{n}
\DeclareSymbolFontAlphabet{\mathcal}{usualmathcal}

\SetArgSty{textnormal}
\SetKwComment{Comment}{{\small\#}~}{}
\SetCommentSty{mycommfont}

\setitemize{itemsep=0pt, parsep=0pt} 				% adjust itemize environment
\setenumerate{itemsep=0pt, parsep=0pt} 				% adjust enumerate environment
\setitemize{itemsep=2pt,topsep=2pt,parsep=0pt,partopsep=0pt,leftmargin=*}
\setenumerate{itemsep=0pt,topsep=2pt,parsep=0pt,partopsep=0pt,labelindent=3pt,leftmargin=*}
\newcolumntype{C}[1]{>{\centering\let\newline\\\arraybackslash\hspace{0pt}}p{#1}}
\newcommand\one{\leavevmode\hbox{\small1\normalsize\kern-.33em1}}
\newcommand{\arXiv}[2][]{%
	\ifthenelse{\equal{#1}{}}%
	{\href{http://arxiv.org/abs/#2}{arXiv:#2}}%
	{\href{http://arxiv.org/abs/#2}{arXiv:#2~[#1]}}}

\def\slashchar#1{\setbox0=\hbox{$#1$}           % set a box for #1
   \dimen0=\wd0                                 % and get its size
   \setbox1=\hbox{/} \dimen1=\wd1               % get size of /
   \ifdim\dimen0>\dimen1                        % #1 is bigger
      \rlap{\hbox to \dimen0{\hfil/\hfil}}      % so center / in box
      #1                                        % and print #1
   \else                                        % / is bigger
      \rlap{\hbox to \dimen1{\hfil$#1$\hfil}}   % so center #1
      /                                         % and print /
   \fi}

\graphicspath{{./figs/}}                % define graphics path

\newcommand\snowmass{\begin{center}\rule[-0.2in]{\hsize}{0.01in}\\\rule{\hsize}{0.01in}\\
\vskip 0.1in Submitted to the  Proceedings of the US Community Study\\ 
on the Future of Particle Physics (Snowmass)
\rule{\hsize}{0.01in}\\\rule[+0.2in]{\hsize}{0.01in} \end{center}}

\makeatletter         
\renewcommand\maketitle{
{\raggedright % Note the extra {
\begin{center}
{\tiny\@author}\\[4ex] 
\@date\\[8ex]
\end{center}}} % Note the extra }
\makeatother

\begin{document}

\begin{center}{\Large \textbf{
CompF3: Machine Learning
}}\end{center}

\begin{center}
Phiala Shanahan,
Kazuhiro Terao,
Daniel Whiteson
(Editors)\\[5pt]

Including contributions from White Paper authors:\\

\author[1,2]{Gert~Aarts}
\author[3]{Andreas Adelmann}
\author[4]{N.~Akchurin}
\author[5,6]{Andrei Alexandru}
\author[7]{Oz Amram}
\author[8]{Anders Andreassen}
\author[9]{Artur Apresyan}
\author[10]{Camille Avestruz}
\author[11]{Rainer Bartoldus}
\author[12]{Keith Bechtol}
% ...

\affil[1]{Swansea University, Swansea SA2 8PP, UK}
\affil[2]{European Centre for Theoretical Studies in Nuclear Physics and Related Areas (ECT*) \& Fondazione Bruno Kessler Strada delle Tabarelle 286, 38123 Villazzano (TN), Italy}
\affil[3]{Paul Scherrer Institute, 5232 Villigen PSI, Switzerland}
\affil[4]{Texas Tech University,  {Lubbock}, TX, 79409, USA}
\affil[5]{The George Washington University, Washington, DC 20052, USA}
\affil[6]{University of Maryland, College Park, MD 20742, USA}
\affil[7]{The Johns Hopkins University, Baltimore, MD 21211, USA}
\affil[8]{Google, Mountain View, CA 94043, USA}
\affil[9]{Fermi National Accelerator Laboratory, Batavia, IL 60510, USA}
\affil[10]{University of Michigan, Ann Arbor, MI 48109, USA}
\affil[11]{SLAC National Accelerator Laboratory, Stanford, CA 94309, USA}
\affil[12]{University of Wisconsin-Madison, 1150 University Avenue Madison, WI 53706-1390}
% ...

{\let\newpage\relax\maketitle}

\end{center}

\section*{Abstract}
The rapidly-developing intersection of machine learning (ML) with high-energy physics (HEP) presents both opportunities and challenges to our community. Far beyond applications of standard ML tools to HEP problems, genuinely new and potentially revolutionary approaches are being developed by a generation of talent literate in both fields. There is an urgent need to support the needs of the interdisciplinary community driving these developments, including funding dedicated research at the intersection of the two fields, investing in high-performance computing at universities and tailoring allocation policies to support this work,  developing of community tools and standards, and providing education and  career paths for young researchers attracted by the intellectual vitality of machine learning for high energy physics.
\bigskip\bigskip\bigskip

\snowmass

\clearpage 

\vspace{10pt}
\noindent\rule{\textwidth}{1pt}
\tableofcontents\thispagestyle{fancy}
\noindent\rule{\textwidth}{1pt}
\vspace{10pt}

\clearpage
%%%%%%%%%%%%%%%%%%%%%%%%%%%%%%%%%%%%%%%%%%%%%%%%%%%%%%%%%%%%%%%%%

\section*{Executive Summary}

This report on artificial intelligence (AI) and machine learning (ML) in high-energy physics (HEP) constitutes the first time that this intersection is represented with a dedicated subgroup in the Snowmass Community Planning process. 

AI can be defined as the branch of computer science aimed at mimicking human intelligence, while ML is the subset of AI which uses statistical learning algorithms to build models based on data. While ML has been used in HEP, particularly in experimental applications, for many years, rapid progress over the last decade including the development of deep learning (complex neural-network based algorithms with significant numbers of layers, or `depth') and continuously evolving computing architectures, has led to a revolution in this area. ML is now ubiquitous, and is enabling previously-intractable computational problems to be framed in new ways, and solved, across industry spheres.

 The recent burst of progress in AI and ML is directly impacting HEP, both by super-charging long-standing applications of ML in this context and by creating completely new opportunities and breaking down historical computational bottlenecks. While it is difficult to predict the future of such a rapidly-evolving field, it is clear that ML will play an increasingly important role in HEP and it is critical that the further development of this intersection is nurtured, especially due to its attractiveness for young researchers.  Below, we summarize the outlook and recommendations made in the full report in several areas (with no priority ranking implied by the ordering): interpretability and validation, theory calculations and detector simulations, data reconstruction/analysis, anomaly detection, detector and accelerator design, community tools, standards, resources, and management, and education and engagement. 
 
 In addition to these specific recommendations, a global theme emerges:  it is crucial that the HEP community nurture the source of the innovations which led to these impactful developments, both by funding interdisciplinary exploration and providing the flexible and open-ended computing resources necessary for high-risk high-reward research.

\paragraph{Uncertainty Quantification, Validation and Interpretability}

It is vital that physicists can validate the decisions of ML models and quantify their uncertainty, a goal  made easier if the inner workings of the models are conceptually accessible to physicists. HEP is not alone in this concern, and can benefit from work in the wider community. 
The HEP community should support continued research into interpretable AI and uncertainty quantification (UQ), including  making public benchmark data sets for rigorous testing and comparison of approaches to physically interpretable AI-UQ for physics, and supporting challenges and competitions to create and compare methods of uncertainty quantification, including bias mitigation.

\paragraph{First-principles theory calculations including detector simulations}
ML for first-principles HEP-focused theory calculations and detector simulations is a rapidly-developing endeavour. As this paradigm expands and matures, there is great potential for transformative impact on the scope and precision of theory studies across areas as diverse as cosmology and lattice field theory. Reaching this potential will likely require the scaling of complex coupled workflows of data generation and training with diverse structures, features, and requirements, to large-scale high-performance computing resources. To achieve this, it will be important to support large-scale allocations for high-risk high-reward exploratory work.

\paragraph{Data reconstruction and analysis} 
Physics-specific ML development is driving the AI-based advancement of the frontiers of HEP research. For example, ML models with physics laws baked in can preserve important aspects of nature, such as equivariance over certain symmetry groups, and contribue in key ways to the interpretability of a model. Technical development in this area of research cannot be successful without key support including for common software tools development, support for technical staff, development of a interdicplinary research community, the development of standardized metrics and and benchmark datasets, and finally close connection between the funding agencies and the research community as this field continues to evolve.

\paragraph{Anomaly Detection}
Anomaly detection is a vital component of the search for new physics at colliders and elsewhere. ML provides a set of  statistical tools for identifying anomalies, but no single strategy has emerged as the most universally powerful, and important challenges remain.
Like other ML efforts, anomaly detection will require significant computing resources, but perhaps more critical is continued community research and development to mature these methods.  Coordinated efforts such as the LHC Olympics 2020 should be encouraged and repeated. The methods developed there will be applicable in both future colliders as well as in non-collider contexts.

\paragraph{Detector and Accelerator design and operation}
 ML tools could reduce accelerator design time and the associated costs significantly whilst simultaneously improving the resulting designs by covering a larger phase-space and yielding a deeper understanding of trade-offs. Furthermore, optimization methods developed with ML techniques can be applied to the operations of experimental facilities including the control, calibration, and monitoring of dynamic systems. Given the push toward the next generation of large HEP experiments, as well as a broad range of smaller experiments under development, automated design and operation methods and tools that can be shared across the discipline of science is both timely and potentially massively impactful. Development in this program requires support that combines research efforts by physicists, engineers, and computer scientists to bring novel ML techniques into the front lines of physics experiment. In addition to salary of non-physicists such as engineers and software developers, the costs of developing custom hardware to bring ML applications into the edge of experimental data taking pipeline must be considered.

\paragraph{Community tools, standards, resources and management}
ML will require significant computing resources, in hardware, software, education, and personnel. Exploratory research should receive healthy support, alongside development and deployment.
In hardware, CPU-based computing will be sufficient for some cases, but others scenarios require the use of specialized processors, such as GPU, FPGAs, or ASICs.  It will be important to not only expand the availability of computational resources to address problems at the intersection of physics and ML, but also to re-visit the structure of computing allocations to support high-risk high-reward work on this frontier. Providing flexible, allocation-free computing resources located at universities will support rapid development pipelines, train the next generation of junior scientists, and enable innovation among the junior researchers currently leading work at this intersection. In software, it is essential that the field maintain a close partnership with the broader community, in academia and industry, but that our tools remain open-source to allow for collaboration and development.

\paragraph{Education and engagement}
Given the rapidly-changing landscape of ML in HEP, it is critical to continue to educate our community and develop a workforce with skills to advance this intersection. As a community, we must create career pipelines both inside and outside academia for the junior scientists who are primarily driving innovation at the intersection of ML and HEP. It is imperative that we take a broad perspective to recruit and retain a diverse cross-section of scientists. Capitalizing on industry expertise with cross-disciplinary collaboration will require community assessments of the ethics of such arrangements, as well as the expansion of open data.

\clearpage

\section{Introduction}
\label{sec:intro}

Over the last decade, the use of machine learning (ML) has become ubiquitous across high-energy physics (HEP); for recent reviews, see Refs.~\cite{Guest:2018yhq,Radovic:2018dip,Carleo:2019ptp,Karagiorgi:2021ngt}.  While particle physics has a long history of the use of neural networks and multi-variate analysis, recent bursts of progress in machine learning have had a ripple effect, yielding qualitatively new developments. These include super-charging long-standing applications to exceed the power of expert-designed heuristics, extend the capacity to handle data with very large dimensionality and data sizes, as well as creating completely new opportunities, such as in fast simulation, anomaly detection and theoretical physics.  While ML was only mentioned in passing in the previous community summer study, its rise in importance motivates a comprehensive dedicated study of the role of ML in HEP, and what is needed to support the future of this young but dynamic and impactful area of research.

Both the physics goals and types of ML paradigms which are used to address them are diverse, spanning from the acceleration of exact first-principles theoretical physics calculations based on generative models, through to online triggering or anomaly detection using computer vision tools. From this rich spectrum, several common themes are emerging. In particular, for many applications there are clear needs for interpretability, validation, and uncertainty quantification of ML-based algorithms (Sec.~\ref{sec:interp}) beyond what is typical for other applications. Moreover, physics problems often have specific features, for example in terms of their symmetries, particular data structures, requirements of robustness, reproducibility, or speed of online training, that are not shared by typical industry applications of ML. As a result, off-the-shelf ML tools are often not sufficient, but rather there is a genuine need for the development of custom ML solutions that build in the physics of the application at hand, i.e., "physics-specific ML" (Sec.~\ref{sec:physicsspecificML}). These features present both opportunities and challenges for our community; opportunities for transformative and innovative new approaches developed by a new generation of scientists literate at this intersection of physics, computer science, and high-performance computing, and challenges in meeting the ever-changing and growing needs of the community in this phase of rapid and unpredictable development. {\bf It also highlights the importance of maintaining a vibrant research community in this interdiscplinary field, capable of adapting methods from industry and broader academia, developing entirely novel methods, and finding new applications for existing methods.}

Given the rapid pace of development of ML, and the breadth of ML paradigms that find applications in HEP, it is impossible to accurately predict the impact or role of this class of tools over the next decade. What is clear, however, is that the current trajectory points towards ever more, and more sophisticated, applications of ML in this context. This has several important consequences. In particular, it will be important for access to high-performance computing to expand, and allocation policies to evolve, to suit the diverse workflows of ML applications, many of which may have a `training' phase whose length is difficult to estimate, complicating planning, and concerted efforts must be made to develop community tools and standards for this new paradigm (Sec.~\ref{sec:communitytools}). Moreover, there is an urgent need for the development of educational pipelines; fully exploiting the potential of ML for HEP will require the engagement of experts with different skill-sets,
such as parallel programming and foundational artificial intelligence, to complement the more physics-oriented expertise of our community. This is one component of a larger need to maintain close connections with the wider ML community in academia and industry; physics-specific challenges can spark broader innovation, and physics can benefit from concepts invented for tasks in adjacent fields. Achieving this will require attention to undergraduate and graduate education, but also to the career pathways available for the postdoctoral researchers and other junior scientists who are driving much of the current innovation at this intersection (Sec.~\ref{sec:education}).

The goal of this document is both to summarize the major HEP applications in which ML-based algorithms have already had or promise impact, and to sketch, as much as possible, the future role of this class of tools to anticipate needs in funding, resources and planning.

\begin{comment}
\begin{itemize}
    \item ML has been very important in the last 10 years. Cite reviews:~\cite{Guest:2018yhq,Radovic:2018dip,Carleo:2019ptp,Karagiorgi:2021ngt}
    \item Some themes have emerged, where ML has contributed to to HEP: first principle calculations, analyzing high-dimensional data, anomaly detection, speeding up simulations, etc.
    \item Given the rapid pace of innovation and the large community of ML research that can find applications in HEP, it's impossible to accurately predict the role, but clearly, ML will be important in the next 10 years.
    \item Goal of this document: describe major areas in which ML has contributed and sketch out possible future role to anticipate needs in funding, resources and planning.
    \item Outline topics discussed.
\end{itemize}
\end{comment}

\section{Uncertainty Quantification, Validation and Interpretability}
\label{sec:interp}

A central goal of high-energy physics is to understand the nature of our Universe. That is, it seeks to do more than simply {\it reveal} the fundamental building blocks of matter and its interactions, but to provide some explanation as to {\it why} it works one way and not some other way. So it is natural, therefore, that in employing powerful ML algorithms to help process the high-dimensional and voluminous data produced by high-energy experiments, physicists would seek to go beyond achieving strong performance in a particular task, but to understand {\it how} the problem has been solved, and to ensure that the solution makes sense.

This desire is more than philosophical, as often ML models are trained using samples of simulated data. While HEP benefits from extraordinarily high-fidelity simulations and large datasets, discrepancies between the data and its simulation-based model do remain, and can lead to sources of bias when an ML model trained on simulation is applied to data.  This is true of any simulation-derived analysis technique, but ML models' ability to extract subtle, non-linear correlations among input features makes them uniquely powerful, but also potentially susceptible to small discrepancies. 
It is therefor vital that ML models be {\it validated}, in which physicists  confirm that the aspects of the simulation on which the model relies are accurately described; one avenue towards validation is to develop models which are {\it interpretable}. Historically, physicists have used heuristic calculations to summarize the information and reduce the dimensionality into a small set of interpretable features, but ML's power to directly analyze high-dimensional datasets and recapture information sacrificed by the high-level summaries make it more difficult to validate and interpret a model.

The need for validation and desire for interpretability are not unique to particle physics, and the community can benefit from the attention paid to it by other fields~\cite{csinterp}.  But the challenge will only grow in importance as  data become more voluminous and high-dimensional, and the field searches for signals of new physics which may leave subtle or rare traces.

\subsection{Interpretable ML}

A major challenge for interpretability of ML models arises precisely from the source of their strength: the ability to non-parametrically describe non-linear functions of high-dimensional inputs. That the effective functional form is not constrained by physical insight allows it to discover unexpected strategies, but also cloaks the learned strategy within a black box.  One can, of course, open the box to examine the specifics of the model's construction, but insight is difficult to extract from thousands (or more) nodes and their millions (or more) connections.

Several approaches have been developed to tackle this important problem.  For deep learning using neural networks with smaller dimensional input spaces, one can compare the performance of the network with and without an input, expand the network function in the basis of an input feature, or  projecting the decision surfaces along physical observables in an effort to gain insight~\cite{Roxlo:2018adx,Chang:2017kvc, 2016arXiv161200410A,Wunsch:2018oxb,Baldi:2014kfa}. Many of these approaches, however, are limited to studying the structure of the model in terms of already-identified physical observables. An extension of this strategy is to assemble a complete basis of interpretable observables~\cite{Komiske:2017aww}, and map the black-box strategy into that space\cite{Faucett:2020vbu}; see example applications to muon~\cite{Collado:2020fwm} or jet substructure identification~\cite{Lu:2022cxg} in collider environments.

An alternative approach is to improve the interpretability of a black box ML algorithm by constraining its internal structure. In the wider ML community, there is extensive study of {\it white-box} algorithms, with a preference for those built from linear, mono-tonic functions, which may trade performable for explainability. These can be seen as providing accurate explanations for approximate models rather than approximate explanations for accurate models.  Another thrust imposes requirements on networks to respect physical symmetries, such as rotational or Lorentz symmetries~\cite{Butter:2022rso,Bogatskiy:2020tje,Bogatskiy:2022hub}, or to insist that the global function be comprised of a restricted set of functions which do not cross theoretical lines such as infrared and collinear safety~\cite{Komiske:2018cqr}. These efforts do not provide explicit explanations for individual model decisions, but global guarantees about the functional form.   Finally, one can attempt to interpret specific model choices by constructing local linear approximations of models~\cite{lime,https://doi.org/10.48550/arxiv.2206.06632}; see early examples in high-energy physics~\cite{Lai:2020byl,Agarwal:2020fpt}.

\subsection{Validation and uncertainty quantification}

The ultimate goal of interpretability for ML models is to ease the validation of their predictions as physically sensible, and to quantify uncertainties that arise from biases they may introduce.  In the end, the the usefulness of physical measurements is tied to the magnitude and reliability of their estimated uncertainties.

Especially important in the case of ML for HEP are sources of {\it systematic} uncertainty, which can arise from several underlying mechanisms. Examples include uncertainty in the modeling of detector response, or lack of knowledge of the value of theoretical parameters which are not of interest. See Refs.~\cite{Nachman:2019dol,Dorigo:2020ldg} for comprehensive reviews, but note that different model, experimental or analysis decisions may lead to variations in observations, but this does not necessarily constitute a source of uncertainty. Equally important is to understand that sources of systematic uncertainty are not created equal; some may be reduced via auxilliary datasets, while others are more descriptive than statistical; see Ref~\cite{Ghosh:2021hrh} for a cautionary tale on how efforts to reduce the dependence of ML on nuisance parameters may only obscure the true uncertainty.

A wide variety of techniques have been developed to quantify uncertainties as propagated through machine learning models. A powerful and straightforward approach is validation of the ML model predictions in data control regions, which could alternatively be used to extract estimates of systematic uncertainties. Recent efforts have been made to develop metrics to quantify uncertainty\cite{Viren:2022qon}.

Nevertheless, the flexibility of ML models allows them to potentially reduce the impact of uncertainties. Early examples of efforts to optimize ML-based analyses to be explicitly robust against uncertainties used neuro-evolution~\cite{CDF:2008dvp} or adversarial models~\cite{Louppe:2016ylz,Shimmin:2017mfk}. An alternative is to condition the network~\cite{Baldi:2016fzo} explicitly on nuisance parameters~\cite{Ghosh:2021roe}, allowing the ML model to adjust to the changing context. Recently, efforts have been made to calculate the gradient of the final result with respect to all analysis parameters, which allows for global optimization and reduction of uncertainties~\cite{DeCastro:2018psv}.  The introduction of differentiable simulations  allow for reducing  data-simulation differences~\cite{Viren:2022qon}, thereby reducing the associated uncertainties.  Another approach to reducing such uncertainties by is to training GANs which can learn to adjust simulation to reduce discrepancies with data~\cite{Viren:2022qon}. Modeling uncertainties are often present even in weakly-supervised or unsupervised learning from data, due to extrapolation from control regions~\cite{Metodiev:2017vrx}.

\subsection{Outlook and recommendations}

Machine learning will continue to play an important role in analysis of HEP data as its complexity grows, both in volume and dimensionality. The community will need to balance interpretability with analysis power to ensure that uncertainties are reliably estimated. Fortunately, in this respect particle physics is not unique, as the wider community has similar concerns, in areas sech as self-driving cars and facial recognition. The statistical demands of HEP applications are however distinct, and so will require careful attention.

To focus attention on this question and develop consensus around how to estimate and report uncertainties in analyses which rely heavily on ML, the HEP-Stats-AI community should create benchmark data sets for rigorous testing and comparison of approaches to physically interpretable AI-UQ for physics~\cite{https://doi.org/10.48550/arxiv.2208.03284}. Additionally, studies which make heavy use of ML should be encouraged to make all code publicly available, to allow for reproducibility. Funding agencies should endorse challenges and competitions to create and compare methods of uncertainty quantification, including bias mitigation. Common AI-UQ methods should be embedded into deep learning software suites, similar to SKLearn, to enable widespread usage, testing, and comparison in the HEP community.
    
\section{Physics-specific ML}
\label{sec:physicsspecificML}

%\subsection{Motivation}
Target applications of ML in HEP often have specific features such as symmetries, invariances, limiting behaviors, or unique data structures, which distinguish these problems from those which appear in other (e.g., industry) contexts. As a result, optimal applications of ML tools for HEP often demands custom, or physics-specific, solutions. This section outlines the diverse types of HEP applications which are likely to be impacted by physics-specific ML methods over the coming years, and highlights the challenges and opportunities that can be anticipated at this intersection.

\begin{comment}
\begin{itemize}
    \item Physics problems (not just HEP) have specific features e.g., symmetries, data structures, etc not shared by industry applications of ML, need to build in physics
    \item spans provably-exact ways of accelerating first-principles physics calculations through to detector simulation
\end{itemize}
\end{comment}

\subsection{First-principles theory calculations including detector simulations}

\begin{comment}
\begin{itemize}
    \item ML not just for experiment or data analysis, also first-principles theory
    \item brief overview of types of applications
    \item Comment about ML incorporated into provably-exact algorithms and outline how this can be done (e.g., accelerate by tuning parameters of known alg, accelerate by changing to easier problem with same solution, accelerate by solving approximately but with known correction)
\end{itemize}
\end{comment}

An emerging and promising application of physics-specific ML is to first-principles theory calculations. This is a diverse and growing area, spanning provably-exact algorithms incorporating ML-based accelerators, though to ML-based proxies for expensive numerical simulations or learned model corrections. 

An important class of theory calculations are those that require exactness, i.e., studies in which there is no room for modeling, approximation, or uncertainty arising from imperfect ML if the rigor of the first-principles framework is to be maintained. Naturally, this requirement places a number of important constraints on the ways in which ML can be employed which are problem-specific and often require custom solutions. One example of this paradigm is the application of ML to lattice quantum field theory calculations, summarized in Ref.~\cite{Boyda:2022nmh}; another is in first-principles simulations for event generation, discussed in Ref.~\cite{Butter:2022rso}. There have recently been promising proof-of-principle applications of ML methods in these contexts spanning generative models for sampling path-integral contributions (in lattice field theory) or phase space (in event generators), through to accelerated observable computation and analysis pipelines, all of which carry exactness guarantees, often achieved through symmetry-preserving ML algorithms and/or the application of a mathematically-rigorous correction step which corrects for imperfect ML at the expense of computational efficiency. %In all cases, poorly trained ML results in slow but correct algorithms, and well-trained ML results in acceleration. 
Efforts to accelerate the computations required for hydrodynamic simulations~\cite{Dvorkin:2022pwo} face similar challenges; each specific application involves different hierarchies of computational scale, requiring different resources, structures (e.g., support for model parallelism) and optimizations (e.g., efficient 4D convolutions).

Another class of applications in the category of first-principles theory studies is efforts to test the consistency of experiments with high-dimensional theory model spaces. For example, it is in many cases straightforward (if computationally expensive) to calculate observables in beyond-Standard-Model theories given a set of theory parameters, while the inverse problem of constraining theory parameters from experimental data is often intractable; the standard approach of scanning over parameters and rejecting those that are not consistent with experimental data scales exponentially in cost with the dimension of the parameter space. This is an application where ML methods are already making an impact, and where the themes of interpretability and validation of Sec.~\ref{sec:interp} are of particular import. For example, various generative ML frameworks have been used to improve the sampling efficiency of searches in high-dimensional supersymmetric parameter spaces, with orders of magnitude of improvement in sampling efficiency~\cite{Hollingsworth:2021sii}. Similar improvements have been achieved in simulation-based inference frameworks~\cite{Morrison:2022vqe}. These efforts have strong parallels with efforts in cosmology, where the goal is to constrain both cosmology and galaxy formation parameters with the highest accuracy from observational data. Proposals have been made to carry out this task using machine learning methods trained using state-of-the-art cosmological hydrodynamic simulations~\cite{Dvorkin:2022pwo}.

\begin{comment}
%\subsection{Detector simulations: Phiala}
\begin{itemize}
    \item Surrogate models, either end-to-end or for pieces of the pipeline ~\cite{Adelmann:2022ozp,Butter:2022rso}~\cite{Butter:2022rso}
    \item Learn corrections to fast approximate models with ML~\cite{Adelmann:2022ozp}
    \item Differentiable programming ~\cite{Adelmann:2022ozp}
    \item Inverse simulations and inference~\cite{Butter:2022rso}
\end{itemize}
\end{comment}

Finally, many of the same challenges arise in the application of ML to first-principles detector simulations, which are essential to link the vast data output of multi-purpose detectors with fundamental theory predictions and interpretation. The computational cost for HEP detector simulation in future experimental facilities will exceed the current available resources; ML-based acceleration thus has an important role to play. Surrogate models or approximations of first-principles simulations based on ML frameworks such as deep generative models have been shown to accelerate simulation pipelines~\cite{Adelmann:2022ozp,Butter:2022rso} either end-to-end, or in key components, while maintaining fidelity.  Of particular value to the physics community are unfolding algorithms~\cite{Andreassen:2019cjw}, which in removing detector effects rather than modeling them, are complementary to the simulation efforts; some inevitable approaches can learn to do both~\cite{Bellagente:2020piv, Howard:2022rhf}. As an alternative to using ML for an entire (component of) a simulation pipeline, ML-based corrections to fast approximate models are finding success~\cite{Adelmann:2022ozp}, as are paradigms based on differentiable programming ~\cite{Adelmann:2022ozp} and inverse simulations and inference~\cite{Butter:2022rso}.

{\bf Outlook and recommendations}

ML for first-principles HEP-focused theory calculations and detector simulations is a rapidly-developing endeavour. As this paradigm expands and matures, there is great potential for transformative impact on the scope and precision of theory studies across areas as diverse as cosmology and lattice field theory. Reaching this potential will likely require the scaling of complex coupled workflows of data generation and training with diverse structures, features, and requirements, to large-scale high-performance computing resources. To achieve this, it will be important to not only expand the availability of computational resources to address problems at this intersection, but also to re-visit the structure of computing allocations to support exploratory and high-risk work on this frontier.

\subsection{Data reconstruction and analysis}
%Modern machine learning methods, in particular deep learning techniques including CNNs, have been widely adapted in HEP data analysis tasks. While they show strong promise in a simple task performance metric in many applications, they are 

Incorporating domain knowledge, also referred to as {\it inductive bias}, for data reconstruction and analysis into a machine learning solution can provide significant benefits including better task performance, better sample efficiency, smaller model size, interpretability and explainability, and robustness against domain shift.  Inductive biases may be based on the specific nature of HEP data,  physics laws, or the requirements and constraints for performing physics analysis using the output of such ML models.

For example, the unique structures of HEP datasets can present unique challenges to ML applications. While convolutional neural networks (CNNs), originally designed for image processing, have been applied to a wide range of HEP data that is either naturally structured as or can be converted into 2D or 3D image formats~\cite{Pumplin:1991kc,Cogan:2014oua,Almeida:2015jua,deOliveira:2015xxd,ATL-PHYS-PUB-2017-017,Lin:2018cin,Komiske:2018oaa,Barnard:2016qma,Komiske:2016rsd,Kasieczka:2017nvn,Macaluso:2018tck,li2020reconstructing,li2020attention,Lee:2019cad,collado2021learning,Du:2020pmp,Filipek:2021qbe,Nguyen:2018ugw,ATL-PHYS-PUB-2019-028,Andrews:2018nwy,Chung:2020ysf,Du:2019civ,Andrews:2021ejw,Pol:2021iqw},
specialized architectures can be designed to address challenges specific to physics data. For instance, HEP image data is often globally sparse yet locally dense (i.e., a small fraction of pixels carry meaningful values but the signal region is densely sampled), which makes it challenging to scale standard CNNs with dense matrix multiplications to large images. In neutrino experiments, CNNs with sparse matrix operations have been used to develop applications that scale to meet the needs of the next generation of experiments with orders of magnitude larger detectors~\cite{Abratenko:2020ocq,Domine:2019zhm}.  There are also challenges associated with multiple modalities of HEP data recorded by multiple distinct type of particle detectors with different geometries. Solutions to such challenges include the development of Graph Neural Networks (GNN) designed to effectively combine different detector information for reconstructing particle flows~\cite{Pata:2021oez}. Future research in this area wil combine techniques specially designed for each detector in an effective manner for multi-modal data analysis that is scalable. 

Another way to incorporate domain knowledge into ML architectures is to incorporate constraints arising from physics laws directly, which mitigates the need to learn such features from training data and hence helps to reduce the model size and complexity while improving training sample efficiency. Furthermore, the resulting models are more interpretable since the physical laws are preserved by design, and are also potentially more generalizable. Examples in this class of HEP specific ML models include QCD-aware deep neural networks~\cite{Louppe:2017ipp,Verma:2021ceh} as well as matrix element calculators~\cite{Maitre:2021uaa}. In particular, equivariant models that preserve symmetry groups have been developed specifically for HEP applications~\cite{cohen2016group,2019arXiv190204615C,Boyda:2020hsi,Favoni:2020reg,Dolan:2020qkr,Bulusu:2021rqz}.  A white paper that summarizes the current state and future directions of research in this regard can be found in~\cite{Bogatskiy:2022hub} with recommendations for a set of metrics to specifically measure the strengths of such ML models and support to develop standardized software tools and develop community talent at this intersection. 

Finally, inspiration to design features of ML architectures specifically for HEP data reconstruction and analysis may come from constraints and requirements for such tasks. For instance, while recurrent neural networks (RNN) have been successfully applied to many examples of sequential HEP data~\cite{Guest:2016iqz,Nguyen:2018ugw,Bols:2020bkb,goto2021development,deLima:2021fwm,ATL-PHYS-PUB-2017-003}, such as lists of particles, the output of RNNs are not permutation invariant. Recently, models including Deep Sets, Transformer, and various types of GNNs have been developed with permutation symmetry in order to address this challenge~\cite{Komiske:2018cqr,Qu:2019gqs,Mikuni:2020wpr,Shlomi:2020ufi,Dolan:2020qkr,Fenton:2020woz,Lee:2020qil,collado2021learning,Mikuni:2021pou,Shmakov:2021qdz,Shimmin:2021pkm,ATL-PHYS-PUB-2020-014}. Another example of such constraints is the need to be robust against nuisance parameters, uncertainties, and issues associated with domain shift. For example, adversarial training can be used to reduce variance of output with respect to nuisance parameters~\cite{Louppe:2016ylz} or avoid dependence on domain bias~\cite{Perdue_2018}. A more generalized approach for an uncertainty-aware models and optimization methods can be found in Ref.~\cite{Ghosh:2021roe}. Finally, development of ML models for object reconstruction is key for improving explainability of physics inference. Examples in this class, including a composite model for an end-to-end data reconstruction, can be found in Refs.~\cite{Drielsma:2021jdv,Pata:2021oez, MicroBooNE:2021nss,Hewes:2021heg,abbasi2021convolutional,Kieseler:2020wcq,Ju:2020xty}. 

{\bf Outlook and recommendations}
Physics-specific ML development has made key contributions in advancing the frontiers of HEP research in a number of ways. ML models with baked-in physics laws can preserve important aspects HEP problems such as equivariance over certain symmetry groups, and provide interpretability of models. Furthermore, by encoding key physics knowledge as a part of a model, it results in quicker learning with less statistics of training sample. In order to boost the R\&D effort in this area, there should be a dedicated support and review criteria that helps research along this category. Such criteria may clearly define metrics concerning the strengths of physics-specific ML (e.g. sample efficiency, learning speed, interpretability). Key ingredients for science-specific ML includes support for both common and application-specific software development, technical research staff,  establishing an interdisciplinary research community as well as a close connection between the funding agencies and the research community.

\subsection{Anomaly Detection}

In addition to searching for hints of new physics motivated by theoretical concepts, there is great value in treating HEP experiments as an exploration of the unknown, and in being prepared for unexpected discoveries.  The power of ML to analyze high-dimensional spaces has already produced a rich literature of its application to the task of {\it anomaly detection}~\cite{Bradshaw:2022qev,Fraser:2021lxm,Ostdiek:2021bem,Atkinson:2021nlt,Finke:2021sdf,Stein:2020rou,Pol:2020weg,Benkendorfer:2020gek,Alexander:2020mbx,Thaprasop:2020mzp,Khosa:2020qrz,Cheng:2020dal,CrispimRomao:2020ucc,Knapp:2020dde,CrispimRomao:2020ejk,Nachman:2020lpy,Andreassen:2020nkr,Hajer:2018kqm,Blance:2019ibf,Cerri:2018anq,Collins:2018epr,Collins:2019jip,DAgnolo:2019vbw,Farina:2018fyg,Heimel:2018mkt,Roy:2019jae,Chekanov:2021pus,Jiang:2022sfw,Hallin_2022}, i.e., the identification of data which are unlikely to be due to known Standard Model (SM) processes.

Various methods of ML-based anomaly detection have already been explored. A major theme is the use of auto-encoders, which map data to a latent space and back, but may fail to similarly map anomalous events.  Another category of efforts attempt to model the density of the SM background in order to identify events which have low likelihood of being drawn from that density, using normalizing flows or kernel methods. Anomaly detection methods may be supervised, or unsupervised, or a hybrid~\cite{Metodiev:2017vrx}.

{\bf Capabilities} The strengths and weaknesses of anomaly detection were extensively studied in the recent LHC Olympics 2020 challenge~\cite{Kasieczka:2021xcg}, which are summarized here.  Several black-box datasets were prepared, in some of which were embedded a variety of new physics signals.  These represent reasonable models of new physics, but cannot span the space of possible signals.  However, they are illuminating as the predictions by the participating teams span a wide range of scenarios, including correctly identified new physics (true positives), incorrect claims of new physics in SM-only datasets (false positives), and missed opportunities (false negatives).

{\bf Outlook and Recommendations} Anomaly detection is a vital component of the search for new physics, at colliders and elsewhere. ML provides a set of  statistical tools for identifying anomalies, but no single strategy has emerged as the most universally powerful.  Important challenges remain, such as confronting data with higher dimensionality, especially in cases where the new physics is non-resonant and so the definition of the background is harder to extract from sidebands. How to respond to a significant detection, and how to quantify the significance of a lack of detection, remain important open questions. In addition, online anomaly detection, in which unusual events are identified at the trigger level, remains an important frontier. 

Like other ML efforts, anomaly detection will require significant computing resources.  Perhaps more critical, however, is continued community research and development to mature these methods.  Coordinated efforts such as the LHC Olympics 2020 should be encouraged and repeated. The methods developed there will be applicable in both future colliders as well as in non-collider contexts.

\subsection{Detector and Accelerator design and operation}
Detector design and optimization for HEP experiments (e.g. design of detector components, accelerator magnets) is an essential and complex task, often involving multi-year processes relying heavily on expert intuition and brute force search methods through the design and layout parameter space. Similar challenges are present in the need to maintain optimal operation of components over time, and to make smart trigger decisions to collect high quality yet unbiased physics data. Advancements made by ML research in predictive modeling and optimization tasks are yet under-explored in this area and have the potential to enable expert-guided, automated tools to design and control experimental instruments and data taking processes.

Within the applied mathematics and optimization communities, a great deal of research has been performed on constrained optimization tasks. The recent rapid progress of AI/ML has accelerated such work and simultaneously enabled new methods. Among the most promising frameworks for approaching  design challenges are Bayesian optimization using Gaussian processes (GPs), reinforcement learning (RL) inspired approaches, surrogate based approaches, and differentiable programming. These approaches are strongly coupled with the tasks in scientific applications, computing capacity, and available models to describe the underlying physics. For example, AI/ML-enabled automated design is already employed in many fields of science, such as in chemistry experimentation~\cite{Shen_2021}, protein and material design, and in ASIC electronics design~\cite{10.1038/s41586-021-03544-w}, all of which have achieved either significant design speed up, improved parameter space optimization, or both. Similarly, in HEP, AI/ML applications are developed for detector modeling and design optimizations that employ differentiable programming frameworks to enable differentiable detector physics modeling as well as generative differentiable surrogates for fast approximation of gradients~\cite{Dorigo:2022gqm,NEURIPS2020_a878dbeb,Diefenbacher:2020rna,Paganini:2017dwg,Paganini:2017hrr}. In addition to design, optimization of control for running facilities including experiment detectors and accelerators is critical for delivering high quality physics data. Optimal control requires calibrations~\cite{Cukierman:2016dkb,Baldi:2020hjm,Cheong:2019upg} and timely diagnostics of a dynamic system~\cite{Scheinker:2022iny} to identify potential issues or predict them in advance.

Integration of powerful ML techniques into the front edge of the experiment's data taking requires innovations in both hardware and software design. Many advancements have been made in the area of {\it Edge-ML} including implementation of ML algorithms on Field Gate Programmable Arrays (FPGAs)~\cite{Govorkova:2021utb,Migliorini:2021fuj,Hong:2021snb,Aarrestad:2021zos,Heintz:2020soy,Iiyama:2020wap,Duarte:2018ite,DiGuglielmo:2020eqx,Summers:2020xiy} and Application-Specific Integrated Circuits (ASICs)~\cite{DiGuglielmo:2021ide}. ML-supported smart physics trigger systems~\cite{Bartoldus:2022zlc} will integrate these research advancements and are critical for the future HEP experiments, including DUNE and HL-LHC, where the rate of data streaming is expected to increased by an order of magnitude or more. These customized software and firmware frameworks are often not only fast bust also energy efficient, and may be used as tools for offline data analysis~\cite{Rankin:2020usv}. 

Finally, there are opportunities for existing ML methods to make immediate impact on the present hardware workflow. For example, a computer vision model may be used for a quality control by identifying defect detector components through a visual scanning~\cite{Akchurin:2022apq}. Applying ML methods to support humans, in particular for repetitive tasks where a constant focus over certain duration of period is difficult, may improve a work quality and reduce risks.

Beyond the automation of individual tasks in designing experiments and operating facilities, there is also an overarching goal of optimizing entire HEP experimental pipelines including physics hypothesis generation, the design and modeling of an experiment to test the hypothesis, experiment construction and data taking, extraction of physics, and finally the return to the first stage of hypothesis generation for the next generation of experiments. This will require a greater scope of discussion and planning including consideration of socio-economic changes and the impact of large projects on society over the lifecycle of an experiment, and it will not be further discussed in this report. However, this is an important research topic to be considered for future funding since the expansion of the scope of automation and optimization can only be expected to accelerate. 

{\bf Outlook and Recommendations}\\
ML tools could reduce design time and the associated costs of HEP experiments significantly, possibly by multiple orders of magnitudes, whilst improving the resulting designs by covering a larger design phase-space and enabling a deeper understanding of design trade-offs. Furthermore, optimization methods developed with ML techniques can be applied to the operations of experimental facilities including control, calibration, and monitoring of dynamic systems. Given the push toward the next generation of large HEP experiments, as well as a broad range of smaller experiments under development, automated design and operation methods and tools that can be shared across science disciplines is both timely and potentially massively impactful. Development of this program requires support for combined research efforts by physicists, engineers, and computer scientists to bring novel ML techniques into the front line of a physics experiment. Development of a set of common tools across experiments and frontiers should be strongly supported in a coherent manner with the projects within each experiment. Such a support should include not only the salary of non-physicists such as engineers and software developers,but also the cost to develop custom hardware to bring ML applications to the front edge of the experimental data taking pipeline.

\section{Community Tools, Standards, Resources and Management}
\label{sec:communitytools}

The previous sections make it clear that ML is emerging as a powerful tool for HEP which can tackle important challenges facing the field as the volume and complexity of experimental data, as well as the volume and complexity of computational theory calculations, grows dramatically. Unfortunately, ML models are often computationally expensive, especially in training, and will require significant resources in terms of providing sufficient computing hardware as well as  dynamic and flexible software which take advantage of industrial efforts but remain adaptable to the specific features of HEP problems. 

\subsection{Current Status and Needs}

\begin{figure}
    \centering
    \includegraphics[height=8cm]{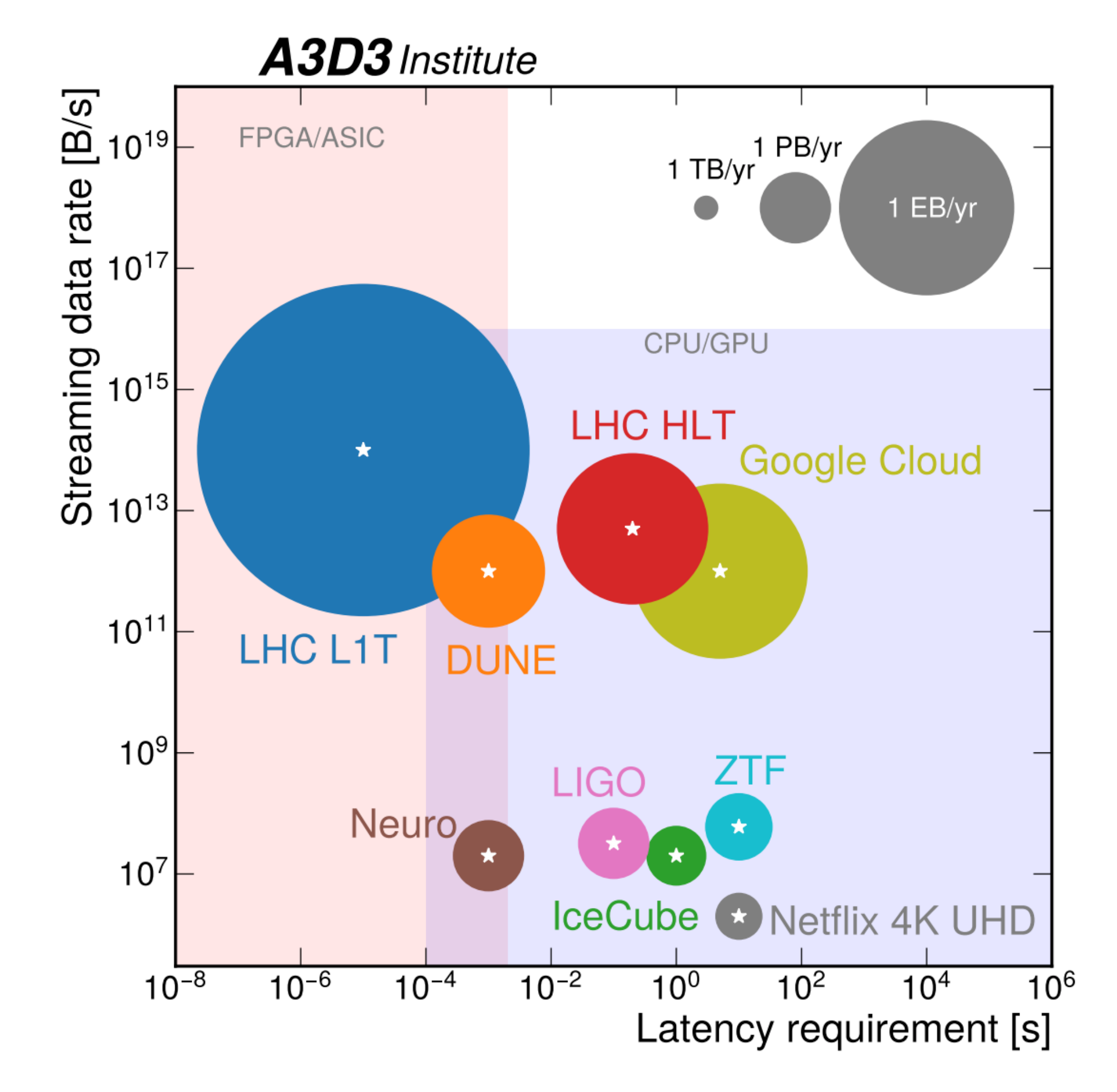}
    \caption{Comparison across various high-energy experiments and industry facilities of the streaming data rate, in units of bytes per second and the latency requirements in seconds.  The area of each bubble indicates the total annual data volume. From~\cite{Harris:2022qtm}}
    \label{fig:compf3latency}
\end{figure}

Machine learning plays an important role in many different aspects of HEP, including collider, neutrino, dark matter, lattice QCD as well as astrophysics, and in many different contexts, including triggering, reconstruction and data analysis.  Figure~\ref{fig:compf3latency} shows a comparison of the data rates versus the latency requirements for some of these systems, demonstrating the extraordinarily wide range of settings.

{\bf  Collider, Neutrino, Astrophysics:} The needs and opportunities of these experiments are explored in detail in Ref~\cite{Harris:2022qtm}, and are summarized here. These communities have fully embraced the use of ML in nearly every aspect of experimental operation, including triggering, data quality monitoring, reconstruction and analysis. These activities span a very wide range of computational needs, including low-latency applications such as triggering and latency-tolerant applications such as offline analysis. With the broadening application of these tools comes a rising computational cost, which will require more than simply additional CPU resources. Use of GPUs has become standard, and special contexts will require FPGAs (for trigger) and ASICs (for radiation-hard environments). The broader use of ML in industry and academia is fueling rapid innovation in hardware, which may soon lead to new technologies such as Tensor PUs, Intelligence PUs and photon-based processing units~\cite{Harris:2022qtm}.  On the other hand, ML may provide some relief for the largest fraction of the computational budget in many experiments, generation of large simulated samples~\cite{HEPSoftwareFoundation:2020daq}.  From a software perspective, a new model of ML resources provides via {\it Software as a Service}, may allow for more flexible deployment of ML resources. Smaller experiments, such as FASER, MicroBooNE, g-2, LUX-ZEPLIN, COHERENT, and DESI also have significant computing needs which should not be neglected~\cite{Andreopoulos:2022iai}.

{\bf Theoretical calculations:} Recent developments in ML-accelerated theory studies represent a new workflow paradigm which requires the availability of computing resources in a pattern distinct from that in experimental work (see Ref~\cite{Boyda:2022nmh} for a review in the context of lattice QCD; similar challenges are faced in the application of ML methods to accelerate cosmological pipelines~\cite{Dvorkin:2022pwo}, first-principles detector simulations~\cite{Hollingsworth:2021sii,Morrison:2022vqe}, and event generation~\cite{Butter:2022rso}). In particular, computational resources need to be made available in a way that is more flexible, responsive, and open-ended, to be compatible with the exploratory and unpredictable nature of ML work in this area, which often involves completely new physics-specific architecture designs. Moreover, resource availability must accommodate the challenges and costs of developing ML algorithms to reach the same scale as state-of-the-art theory calculations via conventional approaches. This will require the availability of both large-scale open-ended allocations on national resources for work at this intersection, and allocation-free university-based resources to enable rapidly-evolving low-overhead developmental work by junior researchers in this area. 
Fully exploiting the opportunities at this intersection will ultimately require development and maintenance of specialized software toolkits~\cite{Boyda:2022nmh}. Just as with software, trained models should be treated as a community resource, particularly for at-scale applications where training may be expensive.

{\bf Direct Detection: } Searches for dark matter in underground or quiet substrates have demonstrated how to improve expected signal-to-background rates as well as particle reconstruction using machine learning methods~\cite{Kahn:2022kae}.  Uncertainty quantification and interpretability are seen as especially important in such low-rate experiments. These experiments will require similar levels of education and training, community standards on uncertainty quantification and interpretability, and access to specialized computing resources~\cite{Roberts:2022ezy}.

\subsection{Outlook and recommendations}

While ML will clearly play a central role in many--if not all--aspects of high energy physics in the next decade, the rapidly-evolving nature of that role makes it challenging to predict what resources and tools will be needed by the community. What is clear is that it will require significant computing resources, in hardware, software, education and personnel, and that exploratory research should receive healthy support, alongside development and deployment.

On the hardware front, CPU-based computing will be sufficient for some use cases, but other scenarios require the use of specialized processors, such as GPU, FPGAs, or ASICs. The field should be prepared to adapt as new hardware is developed which may create transformative opportunities~\cite{Harris:2022qtm}. 

For software, it is essential that the field maintain a close partnership with the broader community, in academia and industry, but that our tools remain open-source to allow for collaboration and development~\cite{campana2022hep,Kahn:2022kae}.

There are efforts within the computer science community to understand the impact of AI research. HEP should participate thoughtfully, as it helps spark innovation and development of AI.

\section{Education and Engagement}
\label{sec:education}

\begin{comment}
\begin{itemize}
    \item Pipeline development: Undergraduate/grad courses at the intersection of DS, ML and physics, involving open software and data used in HEP
    \item Postdoc retention: Important to offer career paths at intersection of ML and physics to retain talented early-career researchers, given overlap with industry careers \cite{Boyda:2022nmh}
    \item Can be difficult to navigate intersection with industry i.e., engagement of those outside our community in our problems, often NDA for collaboration at odds with open science
    \item Innovation-driven developments, primarily by young researchers, can't be concentrated in national labs but must support engagement and development at universities also
    \item Encourage engagement through public data sets and e.g., competitions, examples in Ref. ~\cite{Dvorkin:2022pwo}
\end{itemize}
\end{comment}

Given the continued expansion and rapidly-changing landscape of the applications of ML in HEP, it is clear that education and community engagement in the development of talent at this intersection are crucial and will require careful consideration and investment over the coming years. A particular challenge arises from the hierarchy of expertise in ML across our community; while the current generation of faculty in physics typically do not have formal training in machine learning or even in computer science or high-performance computing, this landscape is changing. Perhaps naturally, much of the transformative work at the intersection of ML and HEP is being undertaken by relatively junior scientists, and the question of how to nurture and retain this talent and what career paths will be open to them within our field is of utmost importance. Looking towards the future, it is imperative to expand the understanding of ML in our community, and that the generation of scientists currently pursuing their degree programs have the opportunity to learn ML in particular as it intersects with HEP and other science domains.

\subsection{Pipeline development}

It is clear that knowledge of not only physics, but of data science in the form of algorithms, high-performance computing, software carpentry, statistics, and ML is becoming a critical tool in HEP. Sustaining innovation at this intersection in our community demands developing and broadening the pool of researchers with this skillset, and integrating computational education within physics curricula. While computer science courses on these topics exist at many institutions, and Massive Open Online Courses (MOOCs), online broadcasting of seminars from interdisciplinary artificial intelligence institutes, and educational-focused journals (e.g., distill.pub), are addressing this need in part in the form of free, online resources, relatively few such resources exist developed by physicists and for physicists that detail the scientifically-informed, data-driven methodologies most relevant to our field. As the HEP/ML intersection continues to develop in scope and complexity, developing such resources as a community will become increasingly important, as emphasized in Ref.~\cite{Benelli:2022sqn}. Simultaneously, it is critical for our community to advocate for and acknowledge the relevance of computational skills for modern physics research. For example, ML for physics must be seen as a legitimate topic for a physics degree, and barriers such as qualifying exams at the graduate level should allow for students specialized in computation. We must make room {\it within physics} for researchers whose work is in the area of physics-informed ML.

In addition to support for developers, physicists interested in {\it applying} ML tools need an opportunity to become familiar with the options, pipelines and pitfalls. There are scattered opportunities at summer schools and one-off training sessions, but much as the modern experiment requires some basic fluency in statistics, particle physics of the present and future needs a basic understanding of machine learning.

As particularly emphasized in Ref.~\cite{Dvorkin:2022pwo}, an important aspect of pipeline development must also be to improve diversity; particularly if innovation is to flourish in our field, the goal must be to nurture critical and questioning perspectives, shaped by a wide breath of experience, particularly experiences different from the traditional norms currently represented in our community. As we engage in the education of the next generation, a broad perspective must be taken to train and engage students in the ethics and efficacy of emerging technologies such as ML. Recruiting and retaining currently minoritized groups is an important piece of expanding the innovation quotient of the workforce.

\subsection{Career paths for junior scientists at the physics/ML intersection}

The current innovation at the intersection of HEP and ML is being driven strongly by junior researchers; to sustain this field, investment must be made in the targeted development of career paths to nurture this talent. In particular, it will be important to make space in our community for permanent positions to retain highly-skilled early-career researchers, particularly those engaged in valuable but technical work which does not lend itself to a traditional academic career trajectory but is readily fostered in industry settings. Increased support for research scientists working on ML for physics will promote a vibrant interdisciplinary community and bridge gaps between different subfields of physics. A number of contributed White Papers detail the challenges faced on this front by particular communities, for example lattice field theory and cosmology~\cite{Boyda:2022nmh,Dvorkin:2022pwo}. 

\subsection{Open data and industry engagement}

The parallel development of ML in HEP and industry contexts creates a wealth of opportunities for cross-engagement. Cross-disciplinary academia-industry collaborations may enable rapid progress and development as well as access to substantial non-traditional funding and computational resources. Fully exploiting such arrangements will require assessing within our community and at a policy level the ethics of such arrangements; for example, non-disclosure agreements for industry collaboration may in some cases be at odds with the principles of open science, but provide valuable technology transfer into our community in the long term. The production and public dissemination of HEP datasets geared towards ML applications, or ML-based coding challenges based on open data, can be a particularly effective tool to create industry engagement~\cite{Dvorkin:2022pwo}. The expansion of such data sharing will require a corresponding expansion of the availability of centralized and decentralized institutional resources (e.g. HPC clusters as well as cloud computing and storage facilities).

\section{Conclusions}

Machine Learning has developed into a powerful set of statistical methods that have influenced nearly every aspect of high-energy physics. While the above sections detail its current role and future outlook along specific thrusts, there are a few global themes which emerge and deserve attention.

{\bf Importance of exploratory interdisciplinary research:} In many cases, ideas which develop into large-scale and impactful veins of research start very small, sparked by a small team of investigators who explore the boundaries between high energy physics and adjacent statistical fields~\cite{Baldi:2014kfa, deOliveira:2015xxd, Louppe:2016aov, Komiske:2019fks, Shimmin:2017mfk, Metodiev_2017,Paganini:2017dwg}. However, very often this research is not directly supported by HEP programs, and relies on individual PIs to cobble together funding from multiple sources for the students and post-docs. The HEP program often supports such ideas once they have gained traction, but care should be taken to also support the source of these innovations, and to ensure that career paths exist for young researchers at the boundaries of HEP and computer science.

{\bf Need for computing support for exploratory research:} Many of the most transformative advances in ML for HEP are arising from novel {\it physics informed AI}. By definition, this does not involve applying known and established ML tools to HEP problems in their standard form, but designing genuinely innovative ML algorithms tailored for HEP problems. Currently, allocation policies at national computing facilities do not support high-risk high-reward exploratory research, where the outcome of significant efforts at training complex ML architectures may be nothing but a deeper understanding of how such architectures behave, which may in the future lead to transformative algorithmic advances. Such exploration is currently concentrated at the dwindling number of universities with local high-performance computing capabilities. Allocation policies at national facilities must be revised to accommodate this important area of research, and new support for hardware at university groups is needed, to fully exploit the potential of ML for HEP.

\begin{comment}
\subsection{Data Science and Machine Learning in Education {\color{red} THIS WP DOES NOT SEEM TO EXIST ON ARXIV}}

The growing role of data science (DS) and machine learning (ML) in high-energy physics (HEP) is well established and appropriate given the complex detectors, large data sets and sophisticated analyses at the heart of HEP research. Moreover, exploiting symmetries inherent in physics data have inspired physics-informed ML as a vibrant sub-field of computer science research. HEP researchers benefit greatly from materials widely available for use in education, training and workforce development. They are also contributing to these materials and providing software to DS/ML-related fields. Increasingly, physics departments are offering courses at the intersection of DS, ML and physics, often using curricula developed by HEP researchers and involving open software and data used in HEP. In this white paper, we explore synergies between HEP research and DS/ML education, discuss opportunities and challenges at this intersection, and propose community activities that will be mutually beneficial.~\cite{}
\end{comment}

%%%%%%%%%%%%%%%%%%%%%%%%%%%%%%%%%%%%%%%%%%%%%%%%%%%%%%%%%%%%%%%%%
\bibliographystyle{SciPost-bibstyle-arxiv}
\bibliography{compf3}
\end{document}